\documentclass[10pt,compsoc]{IEEEtran}
\IEEEoverridecommandlockouts
\usepackage{cite}
\usepackage{amsmath,amssymb,amsfonts}
\usepackage{algorithmic}
\usepackage{graphicx}
\usepackage{textcomp}

\usepackage{booktabs}
\usepackage{color}
\usepackage{url}
\usepackage{multirow}
\usepackage{diagbox}
\usepackage{float}
\usepackage{subfig}
\usepackage{enumitem}
\usepackage{hyperref}

\usepackage[numbers,square]{natbib}

\usepackage[table]{xcolor} 
\definecolor{mycolor}{RGB}{230, 230, 255} 

\def\BibTeX{{\rm B\kern-.05em{\sc i\kern-.025em b}\kern-.08em
    T\kern-.1667em\lower.7ex\hbox{E}\kern-.125emX}}
\begin{document}

\title{Evaluation of Headrest-Integrated Loudspeakers for Enhanced Spatial Audio Immersion in Automotive Cabins\\

}

\author{\IEEEauthorblockN{Martin Wolters, Jacobo Giralt, Harald Mundt, and Arijit Biswas\\}
\IEEEauthorblockA{\textit{Dolby Germany GmbH, Nürnberg, Germany} }



}

\maketitle

\begin{abstract} 
Immersive object-based spatial audio is now firmly established in the music industry as the standard for production, distribution, and playback. The number of automobiles integrating such content to provide premium entertainment experiences is steadily increasing, driving the development of new audio rendering techniques. While loudspeakers integrated into automotive headrests have been around for more than 50 years, they have not yet achieved status as a standard feature in new cars. However, they represent a powerful tool for reproducing immersive audio by enabling the creation of personal sound zones with reduced passenger distraction while effectively complementing existing cabin speakers. We conducted subjective assessments using paired comparison experiments to measure preference and multiple spatial audio attributes. We modeled the resulting probability outcomes using a probabilistic choice model, the Bradley-Terry-Luce rank ordering. The results indicate that headrest-integrated speakers can improve the audio perception in immersive audio scenarios.
\end{abstract}

\begin{IEEEkeywords}
Automotive spatial audio, headrest loudspeakers, binaural rendering, HRTF processing, Bradley-Terry-Luce model
\end{IEEEkeywords}

\section{Introduction}
    
    Loudspeakers mounted to automotive headrests were patented as early as 1970~\cite{patent1970stereo}. Historically, proximity to the listener's ears has been utilized to reduce acoustic energy loss, making headrest speakers an effective solution in high-noise environments such as convertible cars~\cite{motortrend_mx5}. They also facilitate localized audio reproduction for individual passengers while providing a degree of acoustic isolation from other occupants. Bleiholder et al.~\cite{bleiholder2026review} measured level reductions between 7 dB (driver to co-driver) and 13 dB (driver to backseat). This has established headrest-integrated audio as a preferred system for communication-centric use cases. However, more recently, the industry has shifted towards the use of headrest speakers to reproduce immersive audio in the cabin~\cite{brandner20203d, bleiholder2026review}.
    
    Our investigations indicate that the combination of headrest speakers with traditional cabin speakers leads to superior and more versatile sound reproduction. In addition, signal processing techniques can further enhance the quality of playback. In this manner, content intended for reproduction to the side of or behind the listener is routed to the speaker drivers integrated in the headrest instead, or in support of the corresponding cabin loudspeakers. The relevant channels are first processed using binauralization techniques to correct for the intended presentation direction, compensating for the physical placement of the headrest drivers relative to the listener's ears. This approach enables headrest speakers to augment a full 7.1.4 immersive system or a standard 5.1 system to increase spatial envelopment. They can also be combined with discrete front speakers only, reducing sound pressure level for passengers in other seats who may prefer a quieter space. 
    
    Our research evaluates different configurations in a subjective assessment using the attributes of spaciousness, spectral naturalness, loudness, clarity, and overall preference. To analyze the data, we evaluated probability outcomes using a rank-ordering framework, the Bradley-Terry-Luce model~\cite{bradley1952rank}. This class of statistical methodology was successfully employed in previous spatial audio evaluations~\cite{giralt2024enhanced} to establish a robust and mathematically significant hierarchy of preference.  
    
    Our hypothesis is that headrest speakers will become increasingly prevalent in the next generation of automobiles. Tier-1 suppliers~\cite{bose_seatcentric, harman_seatsonic} and OEMs are already creating solutions often focused on communications and personal audio zones. With the results of our study, we quantify and document how headrest speakers optimize immersive audio across different configurations. 

    The remainder of this paper is organized as follows. Section 2 details the headrest audio processing and rendering configurations evaluated in this study. Section 3 outlines the subjective assessment methodology, including the experimental setup, listener task, and the mathematical framework used for data analysis. Section 4 presents the results of the probabilistic choice modeling and statistical analysis. Finally, Section 5 provides a discussion of the perceptual findings, followed by concluding remarks in Section 6.
    
    \section{Headrest Audio Processing and Rendering}

\subsection{Laboratory vehicle and speaker setup}

\begin{figure}[t]
    \centering
    \includegraphics[width=\linewidth]{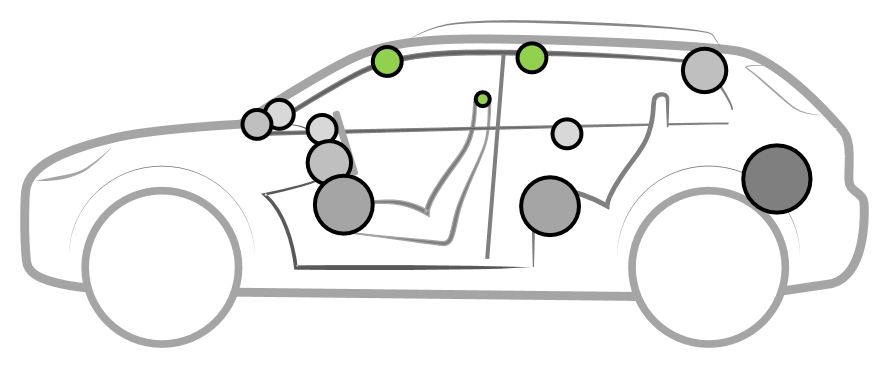}
    \caption{Speaker layout of the lab vehicle from a side view. Circles reflect the approximate size and position of the speakers in the car. Positions highlighted in green are additional to the original system.}
    \label{fig:speaker_layout}
\end{figure}

    The experimental platform is a Volvo XC60 model year 2021 equipped with a premium Bowers \& Wilkins audio system that has been adapted for use as a laboratory vehicle. The system supports multichannel audio reproduction across a variety of speaker configurations, allowing flexible evaluation of different channel layouts and rendering strategies. Each loudspeaker in the cabin can be addressed individually, providing full control over signal routing, level, and timing for each driver independently. In addition to the original speaker system, comprising 15 speaker drivers, the vehicle features two pairs of overhead speakers, and headrests with built-in stereo speakers for the front seats. The resulting speaker layout is depicted in Figure~\ref{fig:speaker_layout}. The headrest-integrated speakers are also individually addressed by two dedicated channels. The headrests are off-the-shelf replacement parts from another vehicle model and are approximately 25 cm wide. The speaker drivers are built inside the headrest facing forward behind a perforated fabric.

\subsection{Rendering configurations}

The three rendering configurations evaluated in this study are designed 
to represent a range of practical automotive deployment scenarios, from 
premium full-cabin immersive systems to cost-optimized personal listening 
zones. All configurations share a common object-based audio rendering 
pipeline that accepts immersive content and produces a channel-based 
output tailored to the active speaker layout. 
The renderer supports a variety of output configurations, including dedicated channels for
headrest-integrated drivers.

\subsubsection*{Configuration 1: Discrete 7.1.4 (Full Cabin)}

The first configuration uses the complete set of cabin loudspeakers 
available in the laboratory vehicle, comprising front, 
surround, and overhead loudspeaker positions. This layout represents the 
reference immersive audio configuration and serves as the baseline 
against which the headrest-augmented systems are compared. Immersive 
object-based content is rendered directly to the physical loudspeaker 
positions, with no headrest drivers active. Cabin tuning, including 
per-channel equalization, delay alignment, and gain calibration, is 
applied across all output channels.

\subsubsection*{Configuration 2: Discrete 7.1.4 + Headrest (Full Cabin with 
Headrest Speakers)}

The second configuration augments the full cabin speaker layout with 
the headrest-integrated drivers. In this configuration, content 
intended for reproduction to the side of or behind the listener is 
routed to the headrest drivers in addition to the 
corresponding cabin loudspeakers. Prior to routing, the relevant 
output channels are processed using binauralization techniques that 
compensate for the physical placement of the headrest drivers relative 
to the listener's ears, correcting for the intended presentation 
direction. The headrest drivers in the vehicle are positioned 
approximately at ear height, with a geometry that corresponds broadly 
to the rear and surround channel directions in a multichannel 
presentation. Low-frequency content from channels assigned to the headrest drivers is mixed into the 
front loudspeakers to compensate for the limited low-frequency 
capability of compact transducers. This configuration is 
intended to evaluate whether headrest speakers can be used to increase spatial 
envelopment and perceived immersion beyond what the full cabin speaker layout 
alone can achieve.

\subsubsection*{Configuration 3: Fronts + Headrest (Reduced Cabin with 
Headrest Speakers)}

The third configuration restricts cabin loudspeaker activity to the front 
loudspeakers only and relies on the headrest drivers to reproduce all 
surround and height spatial content. This layout is representative of 
use cases where acoustic isolation between seating zones is a priority, 
for example, when a rear passenger prefers a quiet environment. The 
absence of rear and overhead cabin loudspeakers means that all spatial 
channels behind or above the listener are rendered exclusively through 
the headrest drivers, again with binauralization applied to preserve 
directional intent. The front loudspeakers handle the content of the front stage 
and the low-frequency reproduction. This configuration evaluates the 
degree to which headrest speakers can substitute for a full surround 
layout while maintaining acceptable spatial quality, and quantifies 
any perceptual trade-offs relative to the full discrete system.

\subsubsection*{Signal Processing Common to Headrest Configurations}

In both headrest configurations, the binauralization stage applies 
direction-dependent filtering to each spatial channel routed to the 
headrest drivers. These filters are derived from head-related transfer 
function (HRTF)~\cite{blauert1996spatial} measurements and are parameterized by the intended 
source direction rather than the physical location of the headrest-integrated driver. This 
compensates for the rear placement of the headrest drivers 
and allows them to perceptually simulate source directions spanning 
the full surround and height sphere. In addition, room-related perceptual cues are introduced through a separate processing stage informed by Binaural Room Impulse Response (BRIR) measurements. This stage contributes the early reflections and reverberation characteristics of the cabin acoustic environment, complementing the directional rendering provided by the HRTF-based filters. 
All configurations are loudness-matched to a common playback level prior to evaluation to ensure that any perceived differences in loudness reflect genuine rendering characteristics rather than gain offsets. Loudness matching is verified objectively by recording the test content using a dummy head placed on the front passenger seat and performing ITU-R BS.1770-4~\cite{itu1770} loudness analysis on the recordings. The analysis reveals a maximum absolute loudness difference of 0.5 LKFS between any pair of systems under test, which can be considered very small.

    \section{Subjective Assessment}
    
    \subsection{Experimental Setup}
\begin{figure} [t]
        \centering
        \includegraphics[width=0.7\linewidth]{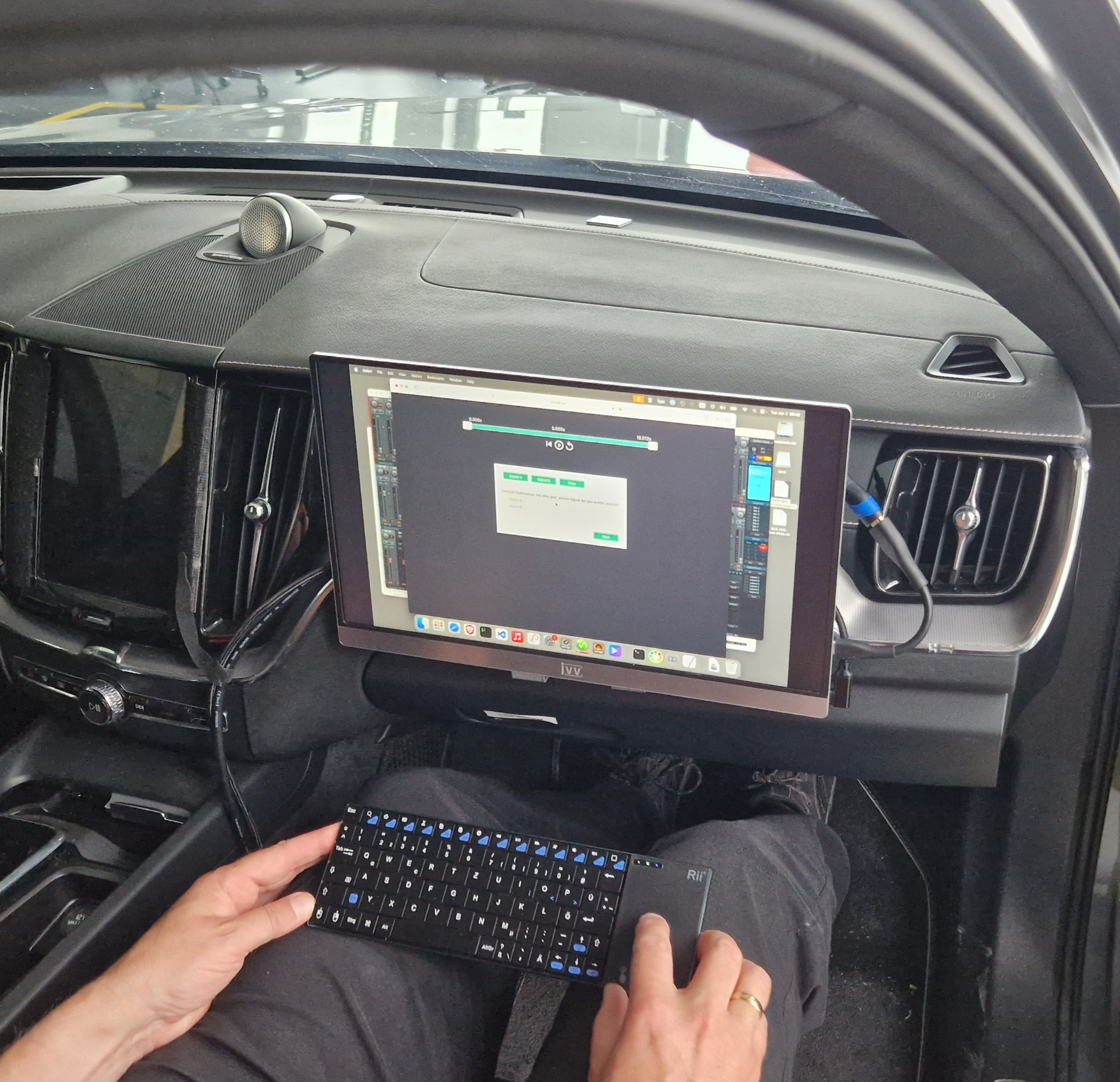}
        \caption{Test setup: Remote screen and wireless keyboard.}
        \label{fig:testsetup}
    \end{figure}

    The listeners were seated in the front passenger seat to make the listening sessions more comfortable (see Figure~\ref{fig:testsetup}). The rendering was fully symmetric and the same results are expected for the driver seat as well. 

    We evaluated six different music excerpts on a panel of listeners ($N=19$). We presented the stimuli in a randomized order to mitigate potential ordering effects. To ensure a consistent understanding of the test framework, all subjects received identical instructions and attribute descriptions prior to the assessment (see Table~\ref{tab:attributes}). During each trial, listeners made a forced-choice paired comparison between two acoustic configurations on five distinct preference questions.

\begin{table}[t]
\scriptsize
\centering
\renewcommand{\arraystretch}{1.15}
\begin{tabular}{p{0.30\linewidth} p{0.62\linewidth}}
\toprule
\textbf{Attribute} & \textbf{Description} \\
\midrule
Overall Preference & 
For this pair, which signal do you prefer overall? \\
\midrule

Loudness & 
For this pair, which signal is louder?

While all of the content is gain matched, you perceive differences in loudness. Please indicate which version ``feels'' louder. \\
\midrule

Spaciousness & 
For this pair, which signal has better spaciousness?

Audio that is ``spacious'' will be well separated and sound like it has width, height, and depth. \\
\midrule

Spectral Naturalness & 
For this pair, which signal is more natural sounding?

Spectral naturalness is another term to describe timbral neutrality. If you are able to detect timbral or color abnormalities that you believe to be outside of the scope of what the content creator intended, this would indicate poor spectral naturalness. \\
\midrule

Clarity & 
For this pair, which signal is clearer?

A signal sounds clearer if it is easier to detect individual instruments, sounds, or voices. Sounds tend to be clearer if they can be separated in the direction they are coming from. \\
\bottomrule
\end{tabular}
\caption{Attributes under evaluation and their description as provided to the listeners.}
\label{tab:attributes}
\end{table}


    During the trials, listeners were instructed to limit their head movements (mirroring typical driver behavior) and were allowed to loop specific audio segments to better isolate discrepancies between the configurations.
    
    \subsection{Data Analysis}

    Evaluating the perceptual tradeoffs of near-field headrest algorithms, particularly balancing spaciousness against spectral naturalness, is a highly complex cognitive task. While absolute grading paradigms can induce listener fatigue when assessing such subtle spatial artifacts, forced-choice paired comparisons provide a highly intuitive and binary assessment framework. To translate these discrete binary choices into a continuous preference hierarchy, we modeled the paired-comparison data using the Bradley-Terry-Luce (BTL) probabilistic choice framework~\cite{bradley1952rank}. The BTL model assumes that for a given perceptual attribute, each audio system $i$ possesses a positive latent worth parameter $w_i$. The probability $P(i \succ j)$ that a listener prefers system $i$ over system $j$ is the ratio of their respective worths:
    
    $$P(i \succ j) = \frac{w_i}{w_i + w_j}$$
    
    We aggregated the individual responses into a global frequency matrix, where $n_{ij}$ represents the total number of times system $i$ defeated system $j$. We then determined the optimal worth parameters by minimizing the negative log-likelihood function $\mathcal{L}(\mathbf{w})$ using the Broyden--Fletcher--Goldfarb--Shanno (BFGS)~\cite{nocedal2006numerical} optimization algorithm:
    
    
    \begin{equation}
    \begin{split}
    \mathcal{L}(\mathbf{w}) = &-\sum_{i=1}^{M-1} \sum_{j=i+1}^{M} \Bigg[ n_{ij} \ln\left(\frac{w_i}{w_i + w_j}\right) \\
    & + n_{ji} \ln\left(\frac{w_j}{w_i + w_j}\right) \Bigg]
    \end{split}
    \end{equation}
    
    We constrained the parameters such that the sum of the probabilities across all $M=3$ systems equals unity ($\sum_{i=1}^M w_i = 1.0$).

    To isolate the specific acoustic characteristics driving listener preference, we computed per-subject Pearson correlation coefficients ($r$) between the binary selection vectors of the individual spatial audio predictors and the overall preference outcomes. Finally, we executed two-tailed binomial tests ($p = 0.5$) on the raw trial tallies to verify the statistical significance of the preference hierarchies.

    To identify underlying demographic trends and prevent nuanced subgroup preferences from being obscured by aggregate averaging, unsupervised machine learning was applied to the listener data. To construct a holistic profile of individual listening preferences, each subject was mathematically represented as a 15-dimensional feature vector. This vector comprises the listener's computed BTL worths for all three evaluated systems across each of the five attributes (3 systems $\times$ 5 attributes = 15 dimensions). We utilized the K-Means algorithm~\cite{macqueen1967kmeans} to partition the subjects into cohesive subgroups exhibiting similar full-test evaluation trends. Principal Component Analysis (PCA)~\cite{jolliffe2002pca} was then employed to orthogonally transform this 15-dimensional space, extracting the principal components of variance and allowing the complex rating trends to be projected onto a two-dimensional visual map. 
    
    Additionally, the relative significance of each dimensional axis is quantified by its \textit{explained variance}. This metric defines the exact percentage of the dataset's total variance, or the total disagreement among the listener panel, that is captured by that specific principal component. Reporting the explained variance provides critical context, mathematically distinguishing whether a divergence in listener preference represents a fundamental demographic split or merely a minor individual quirk. Cluster structure was evaluated using the Silhouette Score~\cite{rousseeuw1987silhouettes}.    
    
    \section{Results and Analysis}
    
    \subsection{Overall Preference and Statistical Validation}
    
\begin{figure}[t]
        \centering
        \includegraphics[width=1.0\columnwidth]{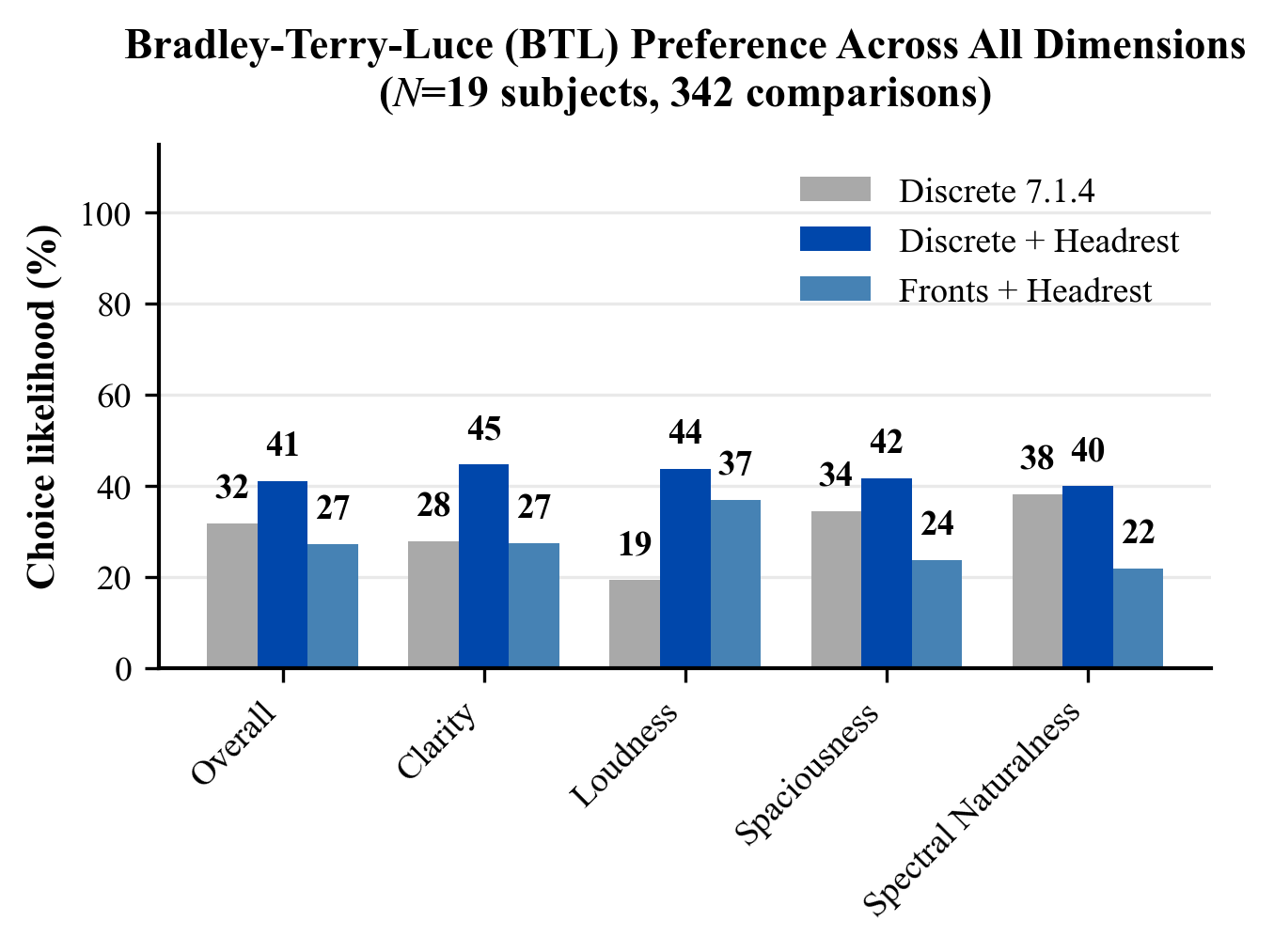}
        \caption{Bradley-Terry-Luce (BTL) probability of preference for the three evaluated spatial audio configurations across Overall Preference, Clarity, Loudness, Spaciousness, and Spectral Naturalness ($N = 19$ listeners). Probabilities are calculated from forced-choice paired comparisons.}
        \label{fig:btl_results}
    \end{figure}

    Figure~\ref{fig:btl_results} presents the BTL probabilistic choice modeling results. The model illustrates the probability of preference across the three evaluated Device Under Test (DUT) configurations: Discrete 7.1.4, Discrete + Headrest, and Fronts + Headrest. 
    
    The BTL probabilistic choice model indicates that the Discrete + Headrest configuration was the most preferred overall, accounting for 41.1\% of the choice probability. Regarding spatial envelopment, Discrete + Headrest achieved the highest BTL worth for Spaciousness (41.7\%).

    To identify significant differences and validate these preference hierarchies, we performed binomial testing across all system combinations ($p \le .05$). Table~\ref{tab:binomial_pvalues} details the resulting $p$-values for all pairwise comparisons across every metric. 

    \begin{table*}[t]
    \centering
    \caption{$p$-values from binomial testing across all attributes and system comparisons. Bold text indicates statistical significance ($p \le .05$).}
    \label{tab:binomial_pvalues}
    \begin{tabular}{@{}lccccc@{}}
        \toprule
        \textbf{System Comparison} & \textbf{Overall} & \textbf{Clarity} & \textbf{Loudness} & \textbf{Spaciousness} & \begin{tabular}{@{}c@{}}\textbf{Spectral} \\ \textbf{Naturalness}\end{tabular} \\ 
        \midrule
        Discrete 7.1.4 vs. Discrete + Headrest      & 0.16 & \textbf{0.02} & \textbf{$<$.001} & 0.40 & 0.93 \\ 
        Discrete 7.1.4 vs. Fronts + Headrest        & 0.40 & 1.00 & \textbf{$<$.001} & 0.07 & \textbf{0.01} \\ 
        Discrete + Headrest vs. Fronts + Headrest   & \textbf{0.05} & \textbf{0.01} & 0.64 & \textbf{$<$.001} &\textbf{$<$.001} \\ 
        \bottomrule
    \end{tabular}
\end{table*}

\begin{figure}[t]
        \centering
        \includegraphics[width=1.0\columnwidth]{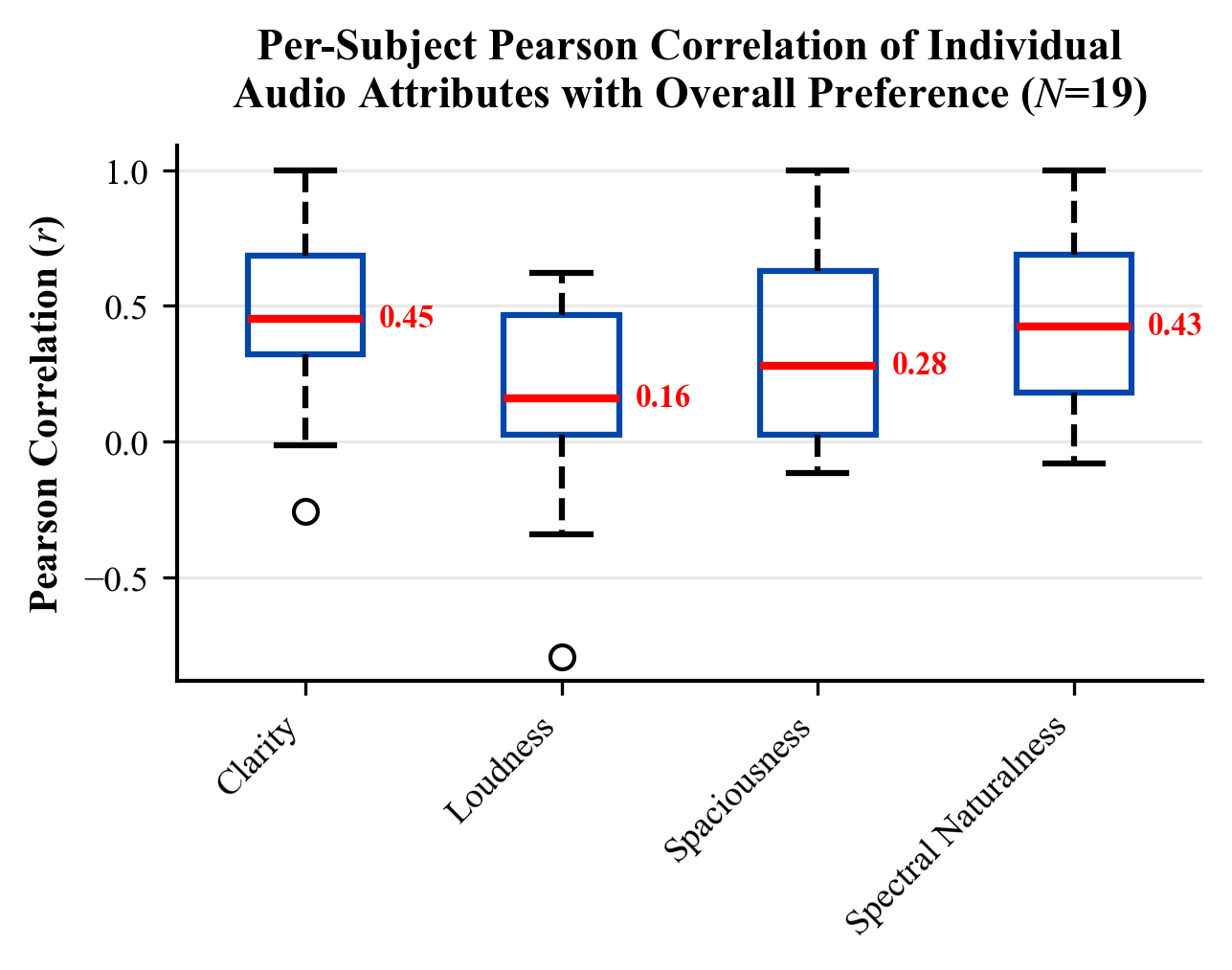}
        \caption{Per-subject Pearson correlation ($r$) of individual spatial audio attributes with Overall Preference ($N = 19$ listeners). Positive correlation values indicate that a preference for a specific attribute heavily drove the listener's overall system choice.}
        \label{fig:correlation}
    \end{figure}

    To understand which perceptual qualities most heavily influenced these overall choices, we mapped per-subject Pearson correlations ($r$), as shown in Figure~\ref{fig:correlation}. Correlation analysis reveals that Clarity exhibited the strongest positive correlation with Overall Preference (median $r = 0.45$). This suggests that clarity was a primary driver of listener preference in this study, aligning with the strong performance of the Discrete + Headrest configuration in that attribute.
    
    \subsection{Content-Dependent Preference}
    
    To investigate how specific source material influenced listener choices, we segmented the BTL probability modeling by the six individual musical excerpts in the test. While full graphical results per excerpt are omitted for brevity, the analysis reveals that the preference ratings are content-dependent. For modern pop and electronic tracks featuring highly discrete spatial objects, Configuration 2 (Discrete + Headrest) dominated across nearly all attributes, specifically excelling in Clarity (capturing up to 58\% of the choice probability). Conversely, for classical and acoustic material, preferences were more distributed; Configuration 3 performed surprisingly well in overall preference, while Configuration 2 maximized spaciousness. Finally, for classic catalog material, the standard Discrete 7.1.4 system remained highly competitive, tying for overall preference and capturing a high BTL worth for spaciousness. This indicates that listeners' preferences vary with genre and mixing style.

    \subsection{Listener Segmentation and Rating Trends}
\begin{figure}[t]
    \centering
    \includegraphics[width=1.0\columnwidth]{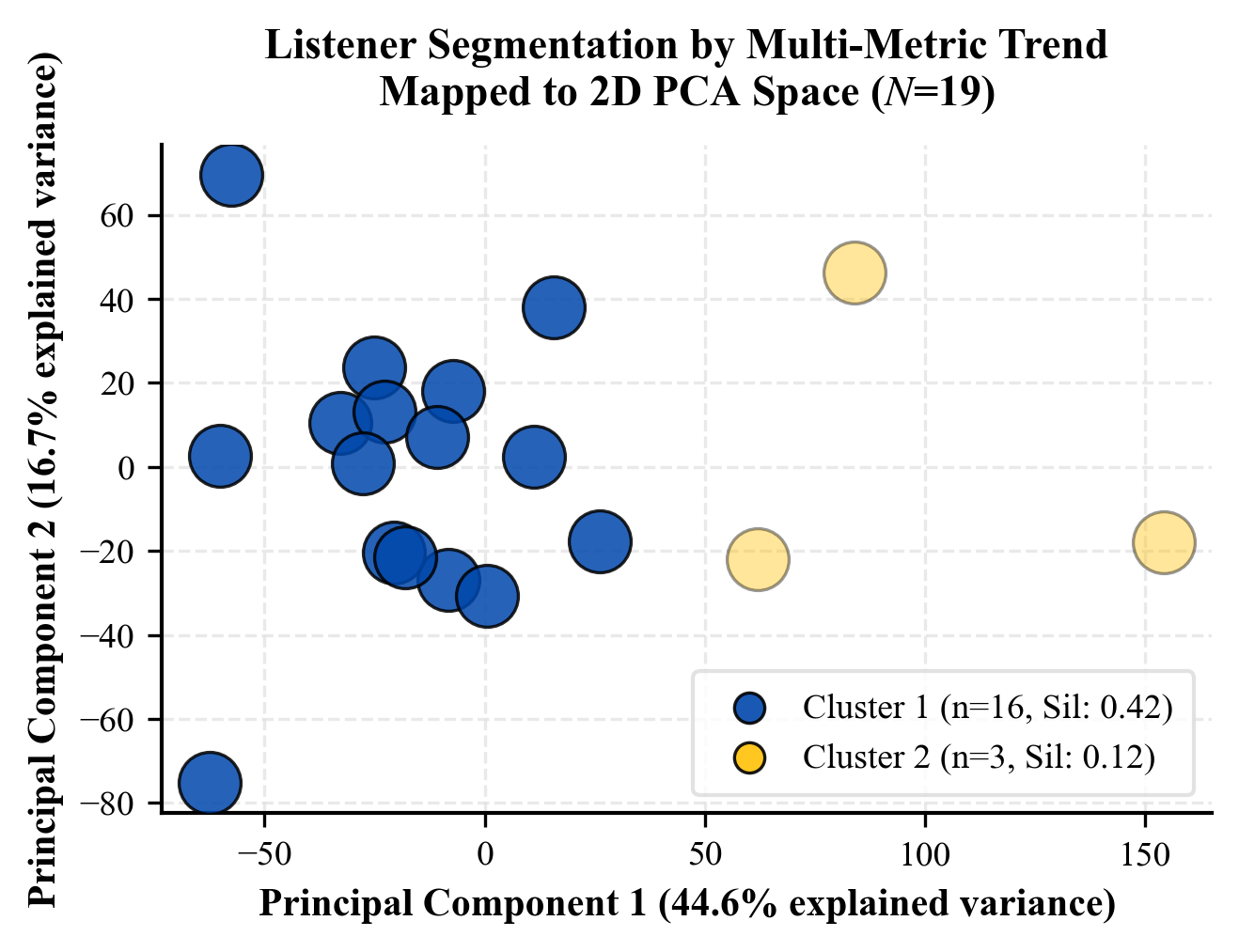}
    \caption{Listener segmentation mapped to a 2D PCA space ($N=19$), with axes indicating the percentage of explained variance. Proximity between markers indicates similarity in individual perceptual profiles, derived from BTL worths across all 15 system-attribute combinations.}
    \label{fig:pca_clusters}
    \end{figure}

    Standard aggregate data analysis assumes a homogeneous listener pool. However, visualizing the K-Means clustering in a 2D PCA space (Figure~\ref{fig:pca_clusters}) reveals distinct perceptual demographics within the panel.

    The PCA diagnostics indicate that Principal Component 1 (the x-axis) is the dominant driver of listener disagreement, capturing 44.6\% of the total dataset variance, which is nearly 2.7 times more than the secondary nuance of Principal Component 2 (16.7\%). 
    
    The cluster analysis successfully identified a highly cohesive main demographic (Cluster 1), representing the vast majority of the panel ($n=16$). This group exhibited a moderately cohesive evaluation trend (Silhouette Score = 0.42). Conversely, a smaller subset of three listeners (Cluster 2) displayed extreme, divergent rating behaviors, pushing them far along the primary axis of variance. 

    Crucially, because these contrarian listeners diverge so heavily from the consensus, the exact mathematical mean of the entire dataset (coordinate [0,0]) lies in an empty region of the map. This indicates that the global mean does not closely correspond to any individual listener, highlighting heterogeneous preference structures within the panel. Instead, the data strongly supports targeting the specific, unified preference profile of the primary cluster.

\begin{figure}[t]
        \centering
        \includegraphics[width=1.0\columnwidth]{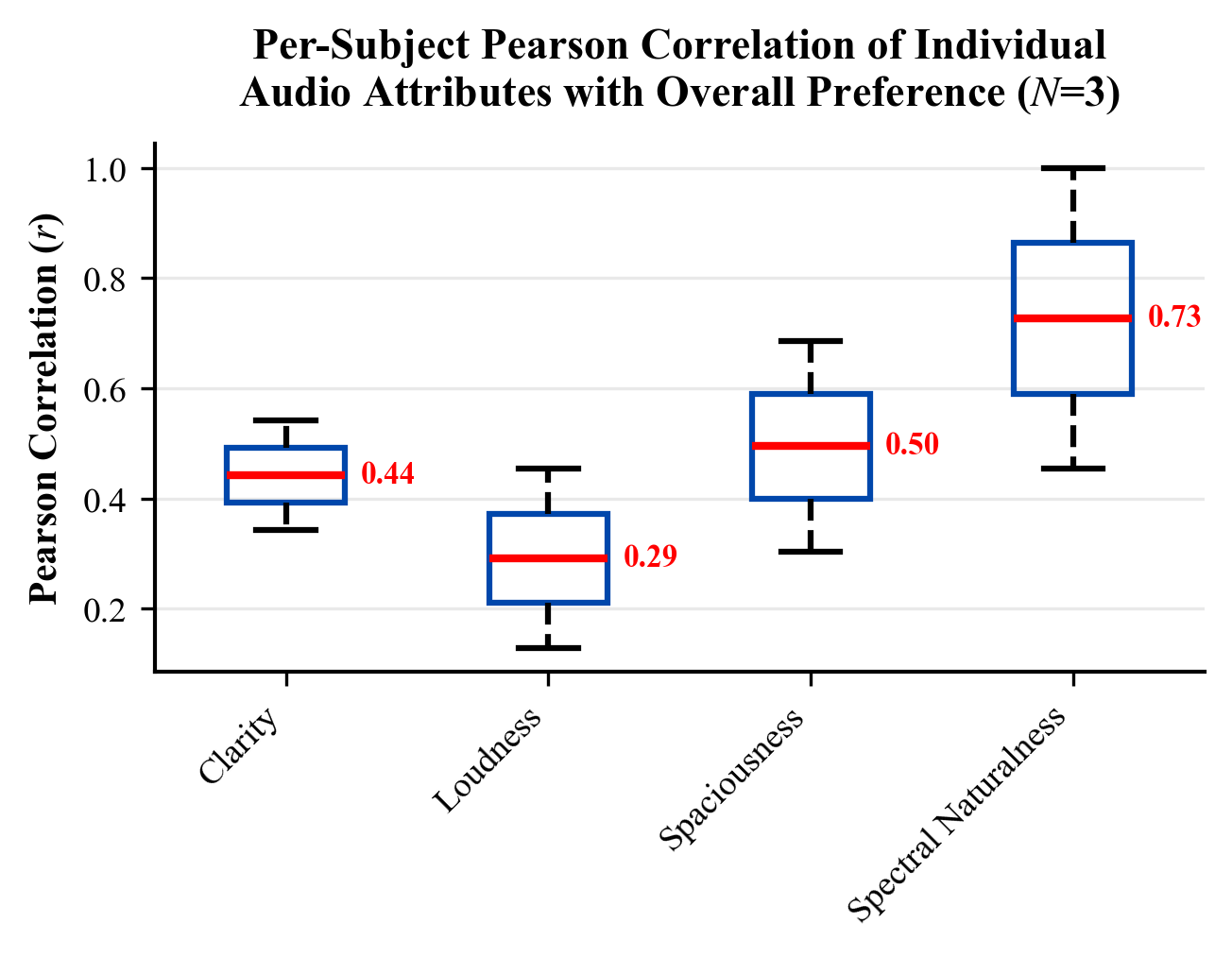}
        \caption{Attribute correlation ($r$) with Overall Preference for the expert subgroup (Cluster 2, $n=3$), dominated by Spectral Naturalness.}
        \label{fig:correlation_cluster2}
    \end{figure}

    Detailed analysis of the listener segments reveals a notable divergence in perceptual criteria. Cluster 1 aligns strongly with the aggregate preference for the headrest-augmented configurations, prioritizing Clarity and Spaciousness. Conversely, Cluster 2 ($n=3$) exhibited a contrary preference, favoring the Discrete 7.1.4 system for Overall Preference. Their data shows a uniquely high correlation with Spectral Naturalness ($r > 0.7$) and lower weighting on Clarity (see Figure~\ref{fig:correlation_cluster2}). These three listeners are engineers engaged in automotive projects, including system tuning and critical listening. While the sample size of this subgroup is small ($n=3$), it presents a preliminary trend suggesting that these expert listeners, potentially influenced by their professional experience, apply more stringent timbral criteria, whereas the broader listener base prioritizes spatial engagement.

    While complete graphical results for the isolated main demographic ($n=16$) are omitted for brevity, recalculating the choice probabilities exclusively for Cluster 1 reveals an even stronger preference for the Discrete + Headrest configuration. This indicates that when the divergent rating criteria of the expert engineering subgroup are removed from the sample, the broader commercial demographic exhibits a heightened preference for the augmented headrest presentation.

    \section{Discussion}
    
Our results show that the listeners clearly preferred Configuration 2 (full cabin with headrest-integrated drivers). This highlights how headrest speakers can significantly enhance immersive audio perception, even in vehicles equipped with high-end audio subsystems. Moreover, headrest speakers offer practical benefits, such as creating personalized zones for phone calls, making them a valuable addition to in-car entertainment. Drivers and passengers will benefit from enhanced audio immersion and greater flexibility.

The results of the listener segmentation lead to additional findings. We noticed that some of the engineers who had previously worked on similar systems evaluated overall preference using different criteria compared to other listeners. For example, while most listeners prioritize ``clarity'' and ``spectral naturalness'', those engineers seem to focus much more on ``spectral naturalness'' and pay very little attention to ``clarity''. This suggests that the way in-car audio systems are tuned should be carefully validated, as the best trade-offs between quality aspects may depend heavily on the target demographic.

Interestingly, Configuration 3 (front speakers only with headrest-integrated drivers) ranked very similarly to Configuration 1 (full cabin) in overall listener preference. This indicates that, in cars with space limitations, such as sports cars with no or limited backseats, a combination of front speakers and headrest-integrated drivers can deliver audio quality comparable to full cabin setups, reducing the need for discrete rear speakers.

The BTL analysis suggests that loudness is not the main driver of preference, allowing us to draw meaningful conclusions. Also, our objective test setup did not lead to large changes in measured loudness between configurations, a finding further supported by listening testers who noted the challenge of determining which version sounded louder. However, listeners more frequently rated configurations with headrest speakers as louder. We attribute this to the close proximity of listeners' ears to those speakers.

During our work, we identified some additional topics for future research. One example is the interaction between different headrest speakers and their optimal tuning. Another example is the comparison of different algorithms in the signal chain, including crosstalk-cancellation and different room models.   

    \section{Conclusion}

With this study, we have documented a commercially available solution that combines a regular full cabin setup with headrest-integrated speakers. We evaluated three configurations using five attributes through a paired comparison listening test, featuring six excerpts from different music genres. Statistical methods, including Bradley-Terry-Luce probabilities (a method for analyzing preference data), were used to interpret the results. Headrest-integrated speakers were preferred by the vast majority of listeners across genres. These findings mark a significant step forward for in-car audio technology, increasing the value proposition for consumers and promoting the wider adoption of headrest speakers in automobiles—more than 50 years after their invention. 
    
     \section{Acknowledgment}
    We would like to thank all of our colleagues in Nürnberg who supported this work by participating in the listening test.

\bibliographystyle{IEEEtran}
\bibliography{IEEEabrv,IEEEexample}

\end{document}